\documentclass[useAMS,usenatbib]{mn2e}
\usepackage{graphicx}
\usepackage{natbib}
\usepackage[varg]{txfonts}

\title[On the dearth of close-in planets around fast rotators]{Secular orbital evolution of planetary systems and the dearth of close-in planets around fast rotators}
\author[A.~F.~Lanza and E.~L.~Shkolnik]{A.~F.~Lanza$^{1}$\thanks{E-mail:
nuccio.lanza@oact.inaf.it} and E.~L.~Shkolnik$^{2}$\thanks{E-mail: shkolnik@lowell.edu}\\
$^{1}$INAF-Osservatorio Astrofisico di Catania, Via S.~Sofia, 78 - 95123 Catania, Italy\\
$^{2}$Lowell Observatory, 1400 West Mars Hill Road, Flagstaff, AZ 86001, USA}
\begin{document}

\date{Accepted ... Received ...; in original form ...}

\pagerange{\pageref{firstpage}--\pageref{lastpage}} \pubyear{2002}

\maketitle

\label{firstpage}

\begin{abstract}
{
Recent analyses of Kepler space telescope data reveal that transiting planets with orbital periods shorter than $2-3$~days are generally observed around late-type stars with rotation periods longer than $\sim 5-10$ days. We investigate different explanations for this phenomenon and favor an interpretation based on  secular perturbations in multi-planet systems on non-resonant orbits. In those systems, the orbital eccentricity of the innermost planet can reach values close to unity through a process of chaotic diffusion of its orbital elements in the  phase space. When the eccentricity of the innermost orbit becomes so high that the periastron  gets closer than $\sim 0.05$~AU, tides shrink and circularize the orbit producing a close-in planet on a timescale $\lesssim 50$~Myr. The probability of high eccentricity excitation and subsequent circularization is estimated and is found to increase with the age of the system.  Thus, we are able to explain the observed statistical correlation between stellar rotation and minimum orbital period of the innermost planet by using the stellar rotation period as a proxy of its age through gyrochronology. Moreover, our model is consistent with the $entire$ observed distributions of the rotation and orbital periods $P_{\rm orb}$ for $3 \lesssim P_{\rm orb} \lesssim 15$~days.
}
\end{abstract}


\section{Introduction}

The Kepler space telescope has observed more than 150,000 stars in a field toward the Cygnus constellation finding more than 3000 exoplanet candidates through the method of transits \citep{Batalhaetal13}. For a subsample of 737 late-type star candidates, \citet{McQuillanetal13} have been able to measure the rotation periods by computing the autocorrelation of the out-of-transit flux modulations induced by surface brightness inhomogeneities (starspots). They report a dearth of close-in planets (orbital period $P_{\rm orb} \la 10$ days) around fast rotating stars (i.e., with a rotation period $P_{\rm rot} \la 3-5$ days).

Using periodogram techniques, \citet{WalkowiczBasri13} measured the rotation periods of $\sim 950$ Kepler planet candidate host stars and confirmed the paucity of close-in planets ($P_{\rm orb}\lesssim 6$~days) around fast-rotating hosts ($P_{\rm rot} \lesssim 6$~days) with the exception of candidates having a radius $R_{\rm p} > 6 R_{\oplus}$ that mostly consist of tidally synchronized systems.  
Large companions may indeed induce tidal synchronization of their host stars thus significantly affecting their rotation regime \citep[e.g., ][]{Bolmontetal12}.

We investigate possible 
explanations for such a scarcity of close-in planets around fast rotating host stars. We favor a scenario based on the secular dynamic evolution of planetary systems in which older stars, that have been slowed down by magnetic braking, are accompanied by close-in planets whose orbits have been shrunk by a combination of dynamical interactions with distant planets and tidal dissipation.  Our model accounts for the observed distribution of the orbital periods of candidate planets in the range $3-15$ days.

\section{Observations}
\label{Observations}
\begin{figure*}
\centering{
\includegraphics[width=10cm,height=16cm,angle=90]{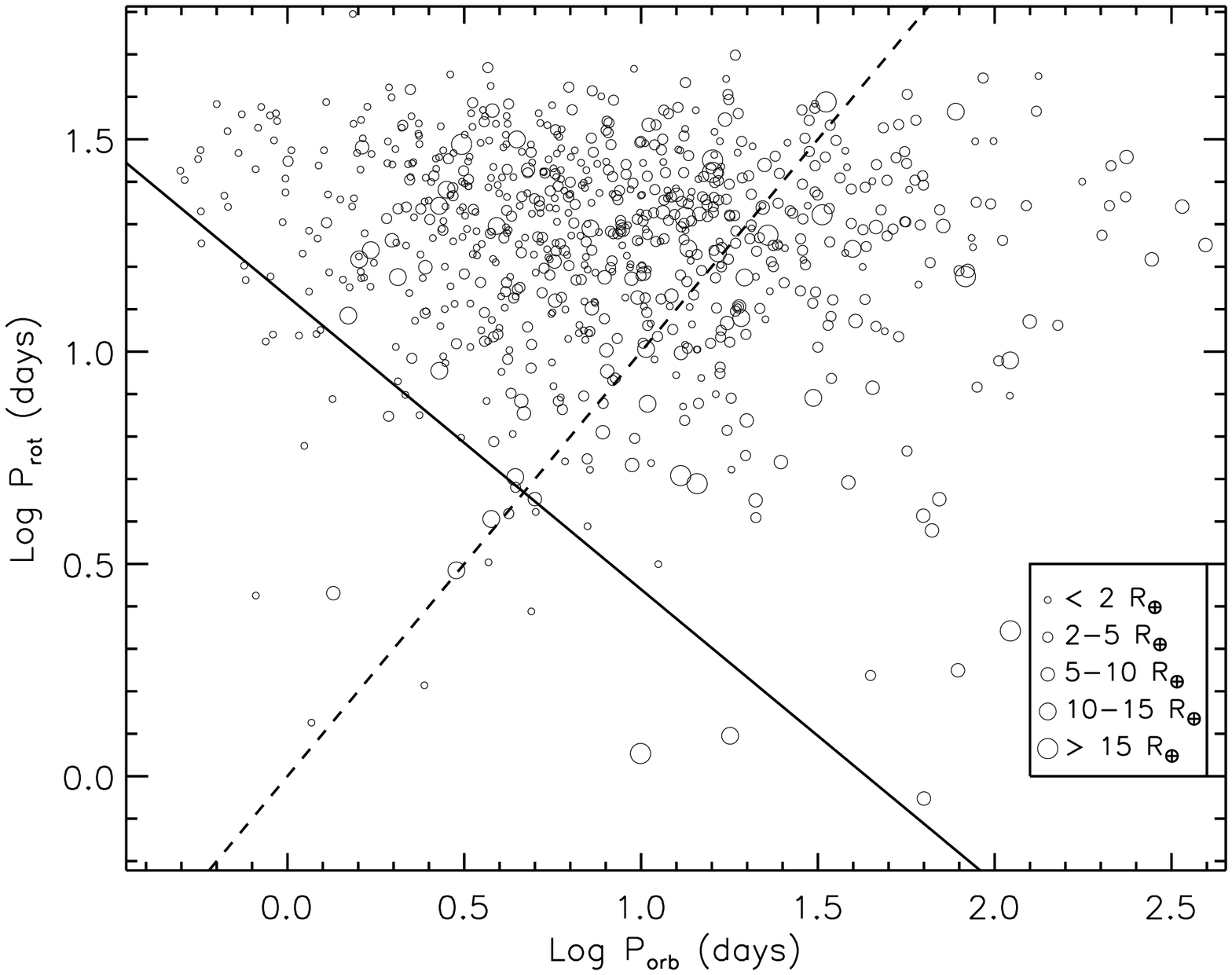}} 
\caption{The rotation period of the host star  $P_{\rm rot}$ vs. the orbital period of its innermost transiting planet $P_{\rm orb}$ for the systems of \citet{McQuillanetal13}. The size of the symbol is proportional to the radius of the planet. The solid line is the regression found by \citet{McQuillanetal13} between the rotation period and the minimum orbital period of the innermost planet. The dashed  line plots the relationship  $P_{\rm rot} = P_{\rm orb}$, i.e., the locus corresponding to tidal equilibrium (synchronization).  }
\label{protvsporb}
\end{figure*}

\citet{McQuillanetal13} reported parameters for a sample of 1961 Kepler objects of interest (KOIs) showing candidate planetary transits. For  737 stars of the sample, they were able to detect the  rotation period by means of the autocorrelation of the photometric time series.  For these stars, they found  a statistical correlation between the rotation period $P_{\rm rot}$ of the host and the minimum orbital period of the innermost transiting planet $P_{\rm orb}$: 
\begin{equation}
\log P_{\rm rot} = (-0.69 \pm 0.07) \log P_{\rm orb} + (1.13 \pm 0.02). 
\label{mcquillan_corr}
\end{equation}
This regression line was computed in order to have 95 percent of the data points in the $P_{\rm rot}$ vs. $P_{\rm orb}$ plot above it in the region bounded by $P_{\rm orb} \leq 10$~days and $P_{\rm rot} \geq 3$~days (see the solid line in Fig.~\ref{protvsporb}). The uncertainties in the fit coefficients were obtained by performing the fit using a random selection of the 80 percent of the data points over one thousand iterations. 
We refer to \citet{McQuillanetal13} for more details on the sample and the criteria applied  to exclude false positives and close stellar binary systems.  Binaries consisting of late-type stars with $P_{\rm orb} \la 10$~days are generally close to  synchronization  (i.e., $P_{\rm rot}=P_{\rm orb}$, plotted as a dashed line in Fig.~\ref{protvsporb}), therefore synchronized binary and planetary systems have been excluded from the fit. 

Most of the planets that follow Eq.~(\ref{mcquillan_corr}) have a radius of $2-3$~R$_{\oplus}$. With few exceptions, their masses have not been measured yet by radial-velocity techniques. Therefore,  we estimate the mass $M_{\rm p}$ of a planet  from its radius $R_{\rm p}$ according to the formula proposed by \citet{Lissaueretal11}: $M_{\rm p} = (R_{\rm p}/R_{\oplus})^{2.06}$ M$_{\oplus}$, where $R_{\oplus}$ is the radius of the Earth and $M_{\oplus}$ its mass.  In this way, we find that most of the planets defining Eq.~(\ref{mcquillan_corr}) have masses ranging between $\sim 4$ and $\sim 10$~M$_{\oplus}$. 

\begin{figure}
\centering{
\includegraphics[width=8cm,height=8cm,angle=90]{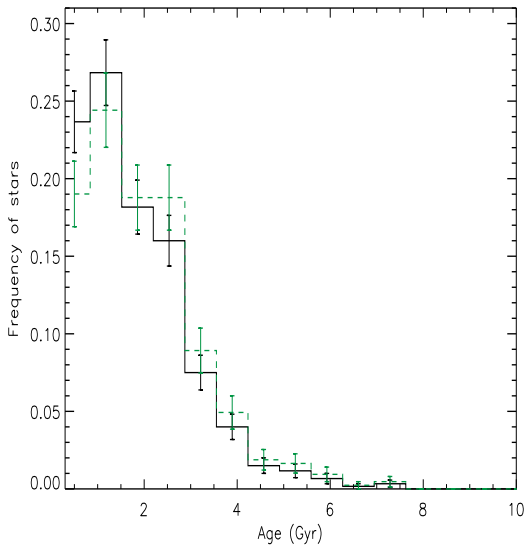}} 
\caption{Distribution of the age of the host stars with measured rotation period, $B-V \geq 0.5$, and $\log g > 4.0$ according to the gyrochronology relationship of \citet{Barnes07} (solid black histogram). The age distribution for the subsample with $0.5 \leq B-V \leq 0.9$, that is less affected by selection effects, is plotted as a green dashed histogram.  Errorbars are given by the square root of the number of stars in each bin.}
\label{age_distr}
\end{figure}

We take advantage of the measured rotation periods to estimate the age distribution of the host stars by means of gyrochronology. We adopt Eq.~(3) of \citet{Barnes07} and determine the $B-V$ color of a given star from its effective temperature  by interpolating in Table~15.7 of \citet{Cox00}. For a proper application of the method, we consider late-type main-sequence stars, i.e., with $B-V \geq 0.5$ and $\log g > 4.0$, having a sample of 707 stars. The distribution of their age is plotted in Fig.~\ref{age_distr} (solid black histogram) where ages younger than 500~Myr have been discarded since Eq.~(3) of \citet{Barnes07}  does not apply to the youngest stars. { Differences arising from the use of different calibrations for the gyrochronology relationship are small, as discussed by \citet{WalkowiczBasri13}.}

Several observational biases affect this distribution. Since Kepler time series show long-term trends and jumps when the spacecraft is re-oriented after $\sim 90$~days of observations, it is not possible to determine reliable rotation periods longer than $\sim 45$~days. This corresponds to an age of $\sim 7.0 \pm 1.5$~Gyr for a star with $B-V = 0.9$ or younger for a  star of later type. 
Assuming a more conservative limit of 30~days as in \citet{Nielsenetal13}, this means a maximum age of  $3.0 \pm 0.7$~Gyr for  stars with $B-V \geq 0.9$. \citet{McQuillanetal14} show that the maximum measured rotation period in the whole sample of Kepler stars increases with decreasing effective temperature ranging from $\sim 20$ to $\sim 50$~days for $T_{\rm eff}$ going from $\sim 6500$ to $\sim 5000$~K. As discussed by the authors, this is likely  associated with the lower amplitude and the lower coherence of the signal for stars of higher effective temperature.  
Therefore, the instrumental limitations and the intrinsic properties of the flux modulation in hot stars severely reduce the number of very old stars in our sample. 

\begin{figure}
\centering{
\includegraphics[width=8cm,height=8cm,angle=90]{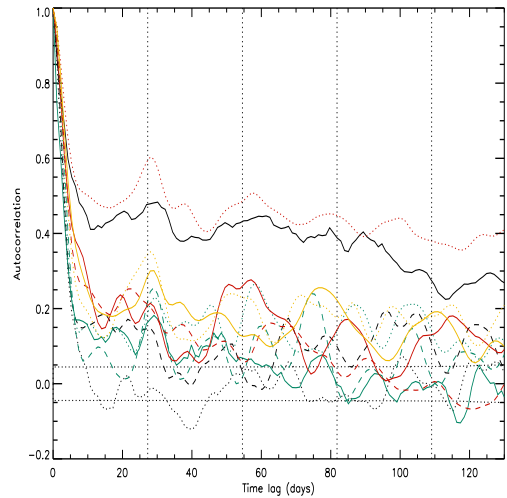}} 
\caption{Autocorrelation of the variation of the total solar irradiance vs. the time lag. Different colours and linestyles refer to the different twelve individual intervals of $\sim 1100$~{days} considered to compute the autocorrelation.  The vertical dotted lines mark the equatorial rotation period of the Sun and its multiples where relative maxima of the autocorrelation are expected to be seen. The horizontal dashed lines mark the autocorrelation level corresponding to noise. }
\label{solar_autocorr}
\end{figure}

To consider a less biased sample, we restrict ourselves to the 478 stars with $0.5 \leq B- V \leq 0.9$ whose gyro-age distribution is plotted as the green dashed histogram in Fig.~\ref{age_distr}. We see that this sample has a more uniform age distribution than the whole sample that extends up to a maximum $B-V$ of 1.6. Nevertheless, even this sample suffers from selection effects. Young stars often show an irregular variability, especially for $T_{\rm eff} > 6000$~K. Moreover, stars with $5000 < T_{\rm eff} < 6500$~K and an age $> 3 -4$~Gyr have active regions that evolve on a timescale significantly shorter than their rotation period making a reliable determination of it particularly difficult. This effect can be investigated by using the 30-yr long time series of the total solar irradiance that is a good proxy for the optical variability of the Sun as a star \citep[e.g., ][]{Lanzaetal03}. In Fig.~\ref{solar_autocorr},  we plot the autocorrelation of the total solar irradiance for twelve individual time intervals of  $\sim 1100$~days that are good proxies for Kepler observations covering  quarters $3-14$ as in  \citet{McQuillanetal13} \footnote{Data obtained from ftp://ftp.pmodwrc.ch/pub/data/irradiance/composite/}. 
The amplitude of the autocorrelation corresponding to the case of  pure noise is indicated by the nearly horizontal dashed lines and has been computed according to \citet{Lanzaetal14}. 
An autocorrelation with multiple peaks equally spaced vs. the time lag, as required for a reliable rotation measurement, is only seen in about $30-40$ percent of the cases. These correspond to periods close to the minimum of the 11-yr solar cycle when the irradiance is modulated by faculae that have lifetimes of $2-3$ rotations. On the other hand, when the modulation is dominated by sunspot groups having a lifetime of only $10-15$ days, it is not possible to measure a clear rotation period \citep[cf. ][ for more details]{Lanzaetal03,Lanzaetal04}. 

We conclude that the hosts  with a measured rotation period are late-type stars with an approximately uniform age distribution up to $\sim 3-4$~Gyr. 
F-type stars have a fraction with measured rotation periods that decreases rapidly with increasing age because of the marked decrease of the amplitude of the modulation with increasing rotation period \citep[see Fig.~4 and discussion in][ for details]{McQuillanetal14}. 
 G-type stars older than $3-4$~Gyr are characterized by a flux modulation too irregular to provide a reliable determination of the rotation period in most of the cases, as demonstrated by the Sun.
On the other hand, K and M-type stars older than about $3-4$~Gyr have rotation periods longer than $\approx 30$~days, i.e., close to or beyond the instrumental limit of Kepler.  In view of these selection effects, host stars with an effective temperature between 5000 and 6100~K without measured rotation periods can be regarded as generally older than those  with measured rotation, i.e.~older than $\approx  4-5$~Gyr.  

\begin{figure}
\centering{
\includegraphics[width=8cm,height=8cm,angle=90]{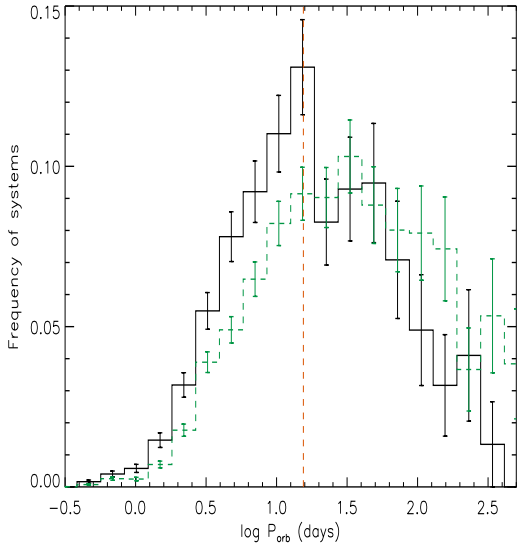}} 
\caption{Distribution of the orbital periods of the planets for the 737 KOI with a measured host rotation period (solid black histogram) and for the  1224 systems for which the rotation period was not detected (dashed green histogram). The distributions have been corrected for the transit probability and then normalized to the total number of systems in each sample, respectively. The vertical dashed orange line marks the beginning of the decrease in frequency of planets around the stars with a measured rotation period observed for $P_{\rm orb} \gtrsim  15$~days. }
\label{porb_distr}
\end{figure}

Having derived some information on the age distribution of the host stars, we now plot the distributions of the orbital periods of the transiting planet candidates   
in Fig.~\ref{porb_distr} both for  stars with and without measured rotation period. The errorbars are given by the square root of the number of candidates in each bin of the histograms. The frequencies have been corrected for the effect of the transit probability $p_{\rm tr}$ by dividing the observed number of candidates in a given bin by $p_{\rm tr} = R_{*}/a$, where $R_{*}$ is the radius of the star and $a$ the semimajor axis of the orbit assumed to be circular. 
The probability that the two distributions come from  the same population is only $9.3 \times 10^{-6}$ according to a Kolmogorov-Smirnov test. 

We see a remarkable decrease in the frequency of candidate planets for $P_{\rm orb} \gtrsim 15$~days around the stars with a measured rotation period, while for those orbiting stars without detected rotation, the decrease, if any, is more gradual. Given the larger errorbars for $P_{\rm orb} > 50$~days, any difference between the two distributions in that interval is not significant. 

In view of the above results  on the age distributions of the host stars, the difference observed in the two distributions for $P_{\rm orb} \lesssim 15$~days is probably due to evolutionary effects. Specifically, we conjecture that the remarkable decrease in frequency of close-in candidate planets with $P_{\rm orb} \lesssim 4-6$~days may indicate their disappearance as a consequence of a complete evaporation or falling into the host star after an evolution of $\sim 4-5$~Gyr. We consider these processes in some detail in the next sections.  

\section{Models}
We consider three different scenarios to account for the correlation found by \citet{McQuillanetal13} and the observations described above,  first discarding hypotheses that {do not appear plausible}, (Sects.~\ref{scen3} and~\ref{scen2}) and conclude with the most likely explanation for the paucity of planets around rapidly rotating stars (Sect.~\ref{scen4}). 

\subsection{Scenario 1: Planet evaporation}
\label{scen3}

The possibility that  planetary evaporation  may account for the observations by completely ablating planets close to very active, rapidly rotating hosts,  
can be quantitatively studied by using  the model by \citet{LecavelierdesEtangs07} to estimate the mass loss rate under the action of the stellar {extreme ultraviolet (EUV)} irradiation.
We derive the mass of our transiting planets from their radius and obtain values up to $\sim 17$~M$_{\oplus}$ for $R_{\rm p} = 4$~R$_{\oplus}$. With an upper limit to the evaporation rate of $10^{12}$~g~s$^{-1}$ for the closest Neptune-like planets (semimajor axis $a \sim 0.03$~AU), we obtain a lower limit of $3.3$~Gyr for their lifetime.  
The evaporation rate is proportional to $a^{-2} R_{\rm p}^{3} M_{\rm p}^{-1} \propto a^{-2} R_{\rm p}$, considering Eq.~(14) of \citet{LecavelierdesEtangs07} and our assumed dependence of the mass on the radius. Thus the planetary lifetime scales as $\sim a^{2} R_{\rm p}^{-1}$, making evaporation ineffective for small planets  ($ 2 \lesssim R_{\rm p} \lesssim 4$~R$_{\oplus}$) beyond $0.05-0.07$~AU, i.e., at orbital periods  $ \gtrsim 4-7$ days around a sun-like star. 
This is confirmed by \citet{EhrenreichDesert11} and \citet{WuLithwick13}, among others. 

\citet{Lanza13} proposed that the energy released by the reconnection of the stellar and planetary magnetic fields can increase the evaporation rate of the planetary atmospheres up to a factor of $\sim 30$. However, this additional source of energy is effective only for $a \lesssim 0.05$~AU, i.e., at orbital periods $\lesssim 4$~days, therefore it cannot modify the above result. 

\begin{figure}
\centering{
\includegraphics[width=8cm,height=8cm,angle=90]{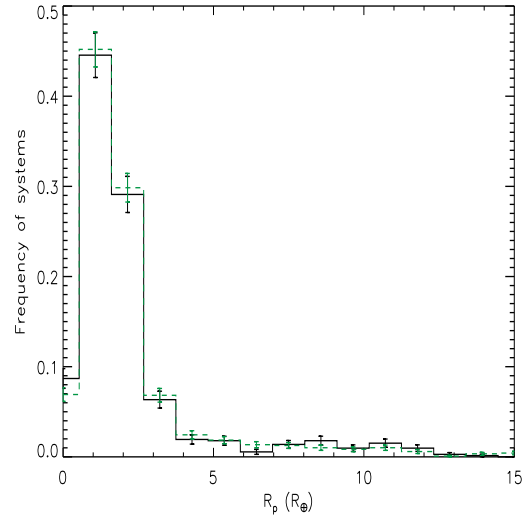}} 
\caption{Distribution of the  planetary radii for the systems whose hosts have measured rotation (solid black histogram) and those without measured rotation (green dashed histogram). The errobars are estimated from the square root of the number of systems in each bin. }
\label{planet_rad_distr}
\end{figure}

We can confirm the limited role of evaporation by comparing the distribution of the planet radii for the systems having hosts with measured rotation with those without in Fig.~\ref{planet_rad_distr}.
Since the systems without measured  rotation are likely to be significantly older than those with measured rotation, their planets should have experienced a stronger cumulative effect of the evaporation, thus making their distribution more peaked toward lower values of the radius. However, we find no indication for that effect because the two distributions are very similar. A Kolmogorv-Smirnov test confirms that they are drawn from the same population with a probability of 0.187. 
We conclude that planetary evaporation is not a viable explanation for the considered correlations that involve planets up to an orbital period of $\sim 15$~days.

\subsection{Scenario 2: Tidal star-planet interactions}
\label{scen2}

Another possibility is that systems with a rapidly  rotating host are the youngest in our distribution and their planets have not had time  to migrate toward the star under the action of tides. On the other hand, stars with longer rotation periods are older and thus there was enough time for the tidal shrinking of the orbits of their planets. Tides remove angular momentum from the orbital motion thus reducing the semimajor axis and spinning up the host star. Nevertheless, { considering planets with masses up to $\approx 10$~M$_{\oplus}$}, this process is  generally not able to modify the stellar rotation period in a significant way, so the only relevant effect is that of decreasing the orbital period. 

More precisely, we can predict the maximum variation of the stellar rotation by considering that the angular momentum of a planet on a circular orbit of semimajor axis $a$ is:
\begin{equation}
\Lambda = \frac{M_{\rm p} M_{*}}{M_{*} + M_{\rm p}} \sqrt{G \, (M_{*} + M_{\rm p}) \, a} \simeq M_{\rm p} \sqrt{GM_{*} a},
\end{equation}
where $G$ is the gravitation constant and $M_{*}$ the mass of the star with $M_{\rm p} \ll M_{*}$. If all this angular momentum is transferred  to a star of radius $R_{*}$ having a moment of inertia $I_{*} = M_{*} (\beta R_{*})^{2}$, where $\beta \simeq 0.36$ is the non-dimensional gyration radius, we obtain an angular velocity variation  $\Delta \Omega_{*} = \Lambda / I_{*}$. For a star with $M_{*} = 1$~$M_{\odot}$,  $R_{*}= 1$~$R_{\odot}$, a maximum planet mass $M_{\rm p} = 10$~$M_{\oplus}$, and maximum orbital separation where tidal interaction becomes effective of $a=0.1$~AU, we obtain: $\Delta \Omega_{*} = 6.7 \times 10^{-7}$~s$^{-1}$ that is only 25 percent of  the angular velocity of a star rotating as slowly as the present Sun. Tidal spin-up occurs on a timescale of at least several hundreds Myr \citep[e.g.][]{Dobbs-Dixonetal04}. { We assume that the stellar convection zone is effectively coupled to the radiative interior, thus the whole star is spun up, not only the convection zone. Indeed, this is consistent with models of the rotational evolution of the Sun and late-type stars \citep[cf. ][]{Bouvier13}.}  Most of the host stars considered by \citet{McQuillanetal13}  have a rotation period shorter than $\sim 30$ days and a mass comparable with that of the Sun, therefore we  conclude that tidal spin-up is ineffective to change their rotation in a significant way, { at least considering planets with masses up to $\approx 10$~M$_{\oplus}$}.  { Note, however, that some authors have considered the possibility of a weak coupling between the convective envelope and the radiative interior in late-type stars \citep[see, e.g., ][]{Winnetal10}. In this case, tides can significantly spin-up the convective envelope because its moment of inertia is at least one order of magnitude smaller than that of the entire star \citep[see, e.g. ][ for the mass of the convection zone in stars of different spectral types]{Pinsonneaultetal01}.  The consequence would be a significant decrease of $P_{\rm rot}$ while the planet approaches the star that would produce a significant fraction of close-in planets around fast-rotating stars, contrary to the observations. Only after the planet in engulfed into the star at the end of its tidal evolution, we would see an absence of close-in planets around fast-rotating stars in agreement with the observations. This would require a tidal decay timescale of the order of a few hundreds Myr, even in the case of planets with an initial orbital period greater than $5-7$ days, in order to keep the number of fast-rotating stars accompanied by close-in planets small. Such a fast tidal evolution is unlikely given the current estimates for the tidal dissipation efficiency in late-type stars (see below). Finally, we note that 
in the case of  F-type stars, internal gravity waves might transfer  angular momentum from the stellar interior to their outer thin convection zone during the main-sequence evolution thus making our simple approach  invalid because the observed rotation period would be modified by this process in addition to the tidal torque \citep[cf. ][ and references therein]{RogersLin13}.}

Next we consider the evolution of a planetary orbit under the action of tides. Assuming that the star is rotating significantly slower than the orbital period as a consequence of  magnetic braking, the evolution of the semimajor axis in the case of a circular orbit is given by \citep[cf., e.g., ][ Eq.~2]{Jacksonetal08}:
\begin{equation}
\frac{1}{a} \frac{da}{dt} = - \frac{9}{2} \sqrt{\frac{G}{M_{*}}} \frac{R_{*}^{5} M_{\rm p}}{Q_{*}^{\prime}} a^{-13/2},
\label{tides-a}
\end{equation}
where $Q^{\prime}_{*}$ is the modified tidal quality factor of the star. It is a non-dimensional measure of the  tidal dissipation inside the star that increases for decreasing $Q^{\prime}_{*}$.  A lower limit to $Q^{\prime}_{*}$ is $10^{6}$ in solar-type stars \citep{OgilvieLin07}. If $Q^{\prime}_{*}=10^{6}$, the radius of the planet $R_{\rm p}$, and the mass of the star $M_{*}$ stay constant, we can integrate Eq.~(\ref{tides-a}) to find the time $\tau_{\rm migr}$ required to migrate to a final semimajor axis $a_{\rm f}$ starting from an initial semimajor axis $ a_{\rm i} \gg a_{\rm f}$:
\begin{eqnarray}
\tau_{\rm migr} &  = & \frac{4}{117} \sqrt{\frac{M_{*}}{G}} \frac{Q^{\prime}_{*}}{R_{*}^{5} M_{\rm p}}  a_{\rm i}^{13/2} \nonumber \\ 
 & \simeq & 7.5 \times 10^{11} \left( \frac{M_{*}}{M_{\odot}} \right)^{1/2} \left( \frac{R_{*}}{R_{\odot}} \right)^{-5} \left ( \frac{M_{\rm p}}{M_{\rm J}} \right)^{-1} \left( \frac{a_{\rm i}}{0.1~ {\rm AU}} \right)^{13/2} \mbox {yr}.
\end{eqnarray} 
Considering an initial orbital period of 4~days for a planet of mass $M_{\rm p}= 10$~M$_{\oplus}$ around a solar-like star, we find $\tau_{\rm migr}  \simeq 7.6$~Gyr that is comparable with the main-sequence lifetime of the star.  Therefore, older stars should have their planets closer than younger stars, if we consider planets with an initial orbital period $P_{\rm orb (i)} \leq 4$~days, while tidal migration takes longer than the main-sequence lifetime of the system for initially longer orbital periods. 

 If we assume a Skumanich-type law of stellar rotation braking, we can relate the present rotation period of the star $P_{\rm rot}$ to its age $t$  as \citep[cf. ][ Eq.~3]{Barnes07}: $P_{\rm rot} \propto t^{n_{\rm g}}$, where the gyrochronology exponent $n_{\rm g} = 0.5189 \pm 0.007 \simeq 1/2$. In the above hypothesis, this time is comparable with $\tau_{\rm migr}$. Therefore, by integrating Eq.~(\ref{tides-a}) and applying Kepler third law, we find the following relationship between the present rotation period of the star, the present orbital period $P_{\rm orb}$ of the innermost planet and its initial orbital period $P_{\rm orb (i)}( > P_{\rm orb}$):
\begin{equation}
P_{\rm rot} \propto (P_{\rm orb (i)}^{13/3} - P_{\rm orb}^{13/3})^{n_{\rm g}} \simeq (P_{\rm orb(i)}^{13/3} - P_{\rm orb}^{13/3})^{1/2}  .
\label{tide_rel}
\end{equation}
Eq.~(\ref{tide_rel}) is not compatible with the dependence $P_{\rm rot} \propto P_{\rm orb}^{-2/3}$ found by \citet{McQuillanetal13}. In other words, tidal evolution alone cannot account for the observations, even considering only the subset of systems with initial orbital periods $\leq 4$~days for which tidal evolution could be relevant. This limit depends on the assumed  lower limit to $Q^{\prime}_{*}$. If $Q^{\prime}_{*} \sim 10^{7}$, as suggested by, e.g., \citet{Jacksonetal09}, the initial orbital period must be $\la 2.5$~days in order to have a significant orbital shrinking during the main-sequence lifetime of the host, thus making our model not applicable to most of the stars that follow the $P_{\rm rot} - P_{\rm orb}$ regression. 

In the above argument, we did not consider in detail the tidal and magnetic braking effects in the evolution of the stellar spin, as done by \citet{TeitlerKonigl14}. However, to reproduce the observed correlation in Eq.~(\ref{mcquillan_corr}), they need to assume $Q^{\prime}_{*}=10^{5}$ that leads to a tidal dissipation rate at least one order of magnitude greater than indicated by the observations of main-sequence close binary stars \citep{OgilvieLin07}. This implies a rapid orbital decay that would strongly deplete the observed population of hot Jupiters \citep[cf., ][]{Jacksonetal09} and should be easily detected with transit timing observations because the difference in the transit mid-times will reach up to several tens of seconds with respect to a constant period ephemeris in $\sim 10$~yr \citep[cf. the case of the very hot Jupiter Kepler-17 in][]{BonomoLanza12}. Moreover, the distribution of planetary systems in the $\log P_{\rm rot} - \log P_{\rm orb}$ plane as simulated by \citet{TeitlerKonigl14} is significantly sparser than the observed one for $P_{\rm orb} \leq 10$~days and $P_{\rm rot} \leq 12$~days. This suggests that a different mechanism is at work to shape the population of close-in planets.

{
We have discussed the tidal evolution by assuming a constant quality factor model because it is widely applied in the literature. However, a constant viscous time model \citep[e.g. ][]{Eggletonetal98} may be more physically motivated. In this case, Eq.~(\ref{tides-a}) becomes:
\begin{equation}
\frac{1}{a} \frac{da}{dt} = -\frac{1}{t_{\rm TF}},
\label{tides-a-ekh}
\end{equation}
where the tidal friction timescale is given by:
\begin{equation}
t_{\rm TF} = \frac{1}{9\sigma_{*}} \frac{a^{8}}{R_{*}^{10}} \frac{M_{*}}{M_{\rm p} (M_{*} + M_{\rm p})} \left( \frac{1-Q_{\rm E}}{Q_{\rm E}} \right)^{2},
\label{tf}
\end{equation}
with $\sigma_{*}$  the stellar tidal dissipation constant and $Q_{\rm E}$ a nondimensional factor of order unity  that depends on the internal stratification of the star and measures its response to the perturbing tidal potential \citep[see Eq.~(40) and Eqs.~(15) and (18c) in ][ respectively]{Eggletonetal98}. Note that we denote the internal structure constant as $Q_{\rm E}$ instead of $Q$ as in \citet{Eggletonetal98}, to avoid any confusion with the tidal quality factor. Repeating the derivation that led to Eq.~(\ref{tide_rel}) starting from Eqs.~(\ref{tides-a-ekh}) and~(\ref{tf}), we find:
\begin{equation}
P_{\rm rot} \propto \left( P_{\rm orb (i)}^{16/3} - P_{\rm orb}^{16/3} \right)^{n_{\rm g}} \simeq \left( P_{\rm orb (i)}^{16/3} - P_{\rm orb}^{16/3} \right)^{1/2}
\end{equation}
that is still incompatible with Eq.~(\ref{mcquillan_corr}), thus showing that the evolution predicted by a viscous tidal model is also not capable to account for the observed correlation. 
}

\subsection{Scenario 3: Secular evolution of planetary systems}
\label{scen4}
Systems consisting of more than two planets are prone to develop a chaotic dynamic evolution as discussed by, e.g., \citet{Laskar96,Laskar08} and \citet{LithwickWu13} in the case of the Solar System. An important consequence is the possibility for the eccentricity of the orbit of  the innermost planet to become very close to the unity, i.e., the planet  closely approaches the star at periastron (cf. Sect.~\ref{secular_chaos}). In this regime, the orbit is circularized by tides inside the planet, while the orbital angular momentum is conserved leading to a close-in orbit (cf. Sect.~\ref{tidal_circs}). This mechanism has been proposed to explain the formation of hot Jupiters \citep{WuLithwick11}. We apply a similar mechanism to the lower mass planets considered by \citet{McQuillanetal13} and show that it accounts for the observed correlation between stellar age as {indicated} by  $P_{\rm rot}$ and the minimum orbital period of the innermost planet $P_{\rm orb}$  as determined by  secular evolution (Sect.~\ref{sec_evol}).  The same mechanism may account also for the  orbital period distribution of the close-in planets ($P_{\rm orb} \lesssim 15$~days; see Sect.~\ref{orb_period_distr}).

\subsubsection{Chaotic dynamics}
\label{secular_chaos}

A system of planets that are far from orbital mean-motion resonances generally evolves through the effect of secular perturbations. If it consists of only two planets on nearly circular and coplanar orbits, its long-term evolution is well described by  Lagrange-Laplace theory. It describes  the evolution of the Cartesian components of the Laplace vectors of their orbits $h_{j} \equiv e_{j} \sin \omega_{j}$, $k_{j} \equiv e_{j} \cos \omega_{j}$, where $e_{j}$ is the eccentricity and $\omega_{j}$ the argument of periastron of planet $j$, as linear combinations of eigenmodes that oscillate sinusoidally in time with characteristic eigenfrequencies. The orbital periods of the two planets stay constant because their mechanical energies are individually conserved, while secular perturbations only imply exchanges of angular momentum between their orbits \citep[cf., e.g., ][]{MurrayDermott99}.  Only if the orbits have { substantial initial  eccentricities or mutual inclination}, the interaction among the eigenmodes can become non-linear and a chaotic regime may develop.\footnote{{ In the case of a mutual orbital inclination between $40^{\circ}$ and $140^{\circ}$, the Kozai-Lidov mechanism can produce large cyclic oscillations in the eccentricity and the inclination of the inner orbit. If the orbit of the outer planet is eccentric,  oscillations in the eccentricity of the inner orbit up to values close to the unity are favored and there is also the possibility of reversing the direction of its angular momentum both along a cyclic or a chaotic evolution depending on the system parameters \citep{Naozetal11,LithwickNaoz11}. In the case of initial nearly coplanar eccentric orbits, the eccentricity of the inner orbit can be cyclically excited to high values  and the orbit may also flip the direction of its angular momentum \citep{Lietal14}.}} On the other hand, in systems consisting of three or more planets, a chaotic regime is generally approached after an initial regular phase during which their orbits stay nearly circular and coplanar. 

The chaotic secular evolution of a system of $N$ non-resonant planets is ruled by the conservation of the  mechanical energy of each of their orbits, that leads to constant semimajor axes $a_{j}$ with $j=1, ..., N$, and of the so-called total angular momentum deficit \citep[hereafter AMD,][]{Laskar96,Laskar97,WuLithwick11}. The exchanges of AMD among planets are particularly important for the excitation of a large eccentricity of the orbit of the innermost planet during the secular evolution of the system. Nevertheless, during most of the time, the orbits of the planets show  small eccentricities and inclinations to the invariable plane, i.e. the plane normal to the total angular momentum of the system \citep[cf., e.g., ][]{Laskar08}. The AMD of the $j$-th planet can be defined as \citep{LithwickWu13}:
\begin{equation}
{\rm AMD}_{j} = M_{j} \sqrt{GM_{*} a_{j}} \left( 1 - \sqrt{1- e_{j}^{2}} \cos i_{j} \right),  
\label{AMD_eq}
\end{equation}
where the mass of the planet $M_{j}$ is much smaller than that of the star ($M_{j} \ll M_{*}$). 

The innermost planet is the one that will generally be subject to the largest chaotic excursions in eccentricity and inclination  depending on the total AMD of the system that drives the secular evolution \citep{LithwickWu13}. During most of its evolution, a system experiences a slow diffusion in  phase space, the effect of which is that of changing the eccentricity and the inclination in a random-walk fashion. The values of these two parameters cannot be predicted  owing to the chaotic character of their evolution, yet their probability distributions are well defined over  intervals of several hundreds Myr or a few Gyrs.

The probability $p_{j}(e,t)$ for planet $j$ of having an eccentricity between $e$ and $e+de$ during a time interval between $t$ and $t+dt$ is given by $p_{j}(e,t) = \varphi_{j}(e,t)\, de\, dt$. \citet{Laskar08} provides a detailed discussion of the distributions of the eccentricities of the Solar System planets  during their evolution based on numerical integrations. Specifically, his distribution $f(e, t)$ at time $t$ corresponds to $f(e,t) = \int_{0}^{t} \varphi(e, t^{\prime})\, d t^{\prime}$ in our notation.  

In the case of the eccentricity, a Rice distribution with mean $m$ and standard deviation $\sigma$ is a very good approximation to $f_{j}(e,t)$ over Gyr intervals \citep[see Eqs. (2) and (3) in][]{Laskar08}:
\begin{equation}
f_{j}(e,t) = \frac{e}{\sigma^{2}(t)} \exp \left[ -\frac{e^{2} + m^{2}(t)}{2 \sigma^{2}(t)} \right] I_{0} \left[ \frac{em(t)}{\sigma^{2}(t)}\right],
\label{laskar}
\end{equation}
where $I_{0}$ is the modified Bessel function of the first kind of degree zero. This distribution is that of the modulus $|z|$ of a complex number $z = x + y \sqrt{-1}$, where $x$ and $y$ are real normally distributed random variables with the same standard deviation $\sigma$ and mean $\bar{x}$ and $\bar{y}$, respectively, so that $m = \sqrt{\bar{x}^{2} + \bar{y}^{2}}$. { The symbol $\sigma$ is distinguished from the tidal dissipation constant $\sigma_{\rm *,p}$.} 
 As noted by \citet{Laskar08}, this implies that the Cartesian components of the Laplace vector of the $j$-th planet $h_{j}$ and $k_{j}$ become random variables because of the chaotic diffusion. 

More precisely, the diffusion of the system in phase space produces a random-walk increase of the standard deviation of the eccentricity vs. the time $t$, i.e., $\sigma(t) = \sqrt{b_{0} + b_{1} t}$, while its mean  stays approximately constant $m(t) \simeq m $. Note that Eq.~(\ref{laskar}) is valid only for $\sigma \ll 1$ because $I_{0}$ is always greater than 1 and the probability density function must decay rapidly to zero for $e$ approaching the unity. Therefore, when the system reaches an approximate equipartition of AMD among the planets, $\sigma$ and $m$ must approach constant values. In the case of the Solar System, this regime is approached only after a timescale longer than the main-sequence lifetime of the Sun because the distribution function of Mercury is still evolving after 5~Gyr, while those of the other terrestrial planets have already become almost constant. Nevertheless, in the case of a system geometrically similar to our own Solar System, an argument based on mechanical similarity \citep{LandauLifshitz69} indicates that the timescale $\tau_{\rm s}$ to reach a stationary density distribution scales as $\tau_{\rm s} \propto a_{N}^{3/2}$, where $a_{N}$ is the semimajor axis of the outermost planet. Therefore, we may assume that the more compact planetary systems detected by Kepler {generally} have reached the stationary phase of their chaotic evolution and the probability density function $\varphi_{1}$ of the eccentricity of their innermost planet no longer explicitly depends on time $t$ \citep[see also  ][]{WuLithwick11}. 
 
Considering a sample of planetary systems with similar initial conditions and the same number of planets, the fraction $P_{1}(e \geq e_{0}, t)$  having the innermost planet with an eccentricity greater than or equal to $e_{0}$ at a given time $t$ is given by: $P_{1}(e \geq e_{0}, t) = \int_{0}^{t}  \int_{e_{0}}^{1} \varphi_{1}(e^{\prime} ) \, dt^{\prime}\, de^{\prime}$, where we have made use of the stationarity of the system to eliminate the dependence of $\varphi_{1}$ on the time. In order to account for Eq.~(\ref{mcquillan_corr}), we shall consider the case of a highly eccentric orbit ($e_{0} \sim 1$) that is subsequently circularized by tides  (see Sect.~\ref{tidal_circs}). 
Defining $\epsilon \equiv (1 -e) \ll 1$ , we can develop the density $\varphi_{1}$ in the interval $[e, 1]$ in a Taylor series truncating it to the first order, i.e., $\varphi_{1} (e) \simeq \varphi_{1}(1) + \varphi_{1}^{\prime}(1) (e-1) = \varphi_{1} (1) - \varphi_{1}^{\prime}(1) \epsilon$ in the interval $ 0 < \epsilon \ll \sigma^{2} \ll 1$. The density function $\varphi_{1}$ is decreasing for $e$ approaching  the unity and is zero at $e=1$, implying that $\varphi_{1}(e) \simeq -  \varphi_{1}^{\prime}(1)  \epsilon $ with $\varphi_{1}^{\prime}(1) < 0$.  
 
If we fix a limit for the probability, i.e., consider the fraction $P_{\rm L}$ of all the systems with $e \geq e_{0}$ at the time $t$, we have $P_{1} (e \geq e_{0}, t) = P_{\rm L}$. Considering the above expression for $\varphi_{1}$, we find:
\begin{equation}
\int_{0}^{t} dt^{\prime} \, \int^{\epsilon_{0}}_{0} - \varphi_{1}^{\prime}(1) \epsilon \, d\epsilon  =  - \frac{1}{2} \varphi_{1}^{\prime} (1) \epsilon_{0}^{2} t  = P_{\rm L},  
\label{prob_distr}
\end{equation}
where $\epsilon_{0} \equiv 1-e_{0}$. Eq.~(\ref{prob_distr}) implies that $\epsilon_{0}^{2} t = (1-e_{0})^{2} t$ stays constant during the stationary phase of the chaotic secular evolution of the considered systems. In other words, if we fix the fraction $P_{\rm L}$, the minimum eccentricity of the systems belonging to the above subset with probability $P_{\rm L}$ varies as $(1-e_{0}) \propto t^{-1/2}$, where $t$ is the time elapsed since the stationary state was reached. 

An estimate of $\varphi_{1}^{\prime}(1)$ can be obtained by considering that in the stationary regime $f_{1}(e)$ tends to a Rayleigh distribution, i.e., it is given by Eq.~(\ref{laskar}) with $m \ll \sigma$ (a detailed derivation is provided in the Appendix~\ref{appendix}). By considering the distribution of the eccentricity for systems with Jovian planets having periastron between 0.1 and 10~AU, i.e., those with planets sufficiently massive to obtain reliable measurements of $e$ from the radial velocity orbits,  \citet{WuLithwick11} find $\sigma \simeq 0.25$. Therefore, we obtain: 
\begin{equation}
\varphi_{1}^{\prime} (1) = \left( \frac{1}{\sigma^{2}} - \frac{1}{\sigma^{4}} \right) \exp \left( -\frac{1}{2\sigma^{2}} \right) \simeq -0.081. 
\label{varphi_value}
\end{equation}
Nevertheless, we shall see below that a larger value, i.e., $\sigma=0.35$ giving $\varphi_{1}^{\prime}(1) = -0.987$, allows us a better reproduction of the observed frequencies of planetary systems (cf. Sect.~\ref{sec_evol}). This is in qualitative agreement with the greater eccentricity  expected for less massive planets,  such as those detected by Kepler, because, for a given AMD, the eccentricity increases with decreasing planet mass (cf. Eq.~\ref{AMD_eq}).

\subsubsection{Tidal circularization of planetary orbits}
\label{tidal_circs}

The results of the previous Section do not consider the effects of tides that tend to circularize the orbit of the innermost planet. They become relevant when the periastron gets closer than  $\approx 0.05$~AU in the case of a solar-mass star. Since we consider orbits with eccentricity close to the unity, the tidal model based on constant tidal quality factors for the star and the planet ($Q^{\prime}_{*}$, $Q^{\prime}_{\rm p}$) is no longer rigorously valid and we should use the so-called constant time-lag model \citep[see Sects.~2 and~3 of ][ for details]{Leconteetal10}. In this hypothesis, the evolution of the eccentricity is  ruled by:
\begin{eqnarray}
\frac{1}{e} \frac{de}{dt} & = & 11 \frac{a}{GM_{*}M_{\rm p}} \left\{ K_{\rm p} \left[ \Omega_{e}(e) - \frac{18}{11} N_{e}(e) \right] \right. \nonumber \\
& + & \left. K_{*} \left[ \Omega_{e} (e) \frac{\Omega_{*}}{n} - \frac{18}{11}N_{e}(e)  \right] \right\},  
\label{dedt}
\end{eqnarray}
where:
\begin{equation}
K_{\rm p} = \frac{3}{2} k_{2 \rm p} \Delta t_{\rm p} \left(\frac{G M_{\rm p}^{2}}{R_{\rm p}} \right) \left( \frac{M_{*}}{M_{\rm p}} \right)^{2} \left( \frac{R_{\rm p}}{a} \right)^{6} n^{2},
\end{equation}
with $n \equiv 2\pi/P_{\rm orb}$ being the mean motion of the planet, $k_{2 \rm p}$ its apsidal motion constant, $\Delta t_{\rm p}$ the constant tidal time lag, $\Omega_{*}$ the angular velocity of rotation of the star,  and the expression for $K_{*}$ is obtained by exchanging the index $\rm p$ with $*$;  the functions $\Omega_{e}$ and $N_{e}$ are given by: 
\begin{eqnarray}
\Omega_{e}(e) & = & \frac{1 + (3/2) e^{2} + (1/8) e^{4}}{(1-e^{2})^{5}},  \\
N_{e}(e) & = & \frac{1+(15/4)e^{2} + (15/8) e^{4} + (5/64) e^{6} }{(1-e^{2})^{13/2}}.  
\end{eqnarray}
For simplicity, we assume that the stellar spin is aligned with the total angular momentum and the 
rotation of the planet is synchronized with the orbital motion because the synchronization timescale of the planet does not exceed $\sim 1$~Myr.  
To allow for a comparison with the model based on the modified tidal quality factor $Q^{\prime}$, we assume that $Q^{\prime}_{*, \rm p} = (3/2) ( k_{2*,\rm p} n \Delta t_{*, \rm p})^{-1}$ that is only approximately valid  \citep[cf. ][]{Leconteetal10}, but is useful in view of the large uncertainties in $Q^{\prime}_{*, \rm p}$ (and $\Delta t_{*, \rm p}$) for stars and planets. { Similarly, we can give the approximate relationship between $Q^{\prime}_{\rm *,p}$ and the viscous dissipation constant: $\sigma_{\rm *,p} [Q_{\rm E}/(1-Q_{\rm E}]^{2} \simeq G /(n R_{\rm *, p}^{5} Q^{\prime}_{\rm *,p})$. We obtain Eq.~(\ref{dedt}) also in the framework of the constant viscous time model provided that $\Delta t_{\rm *, p} = (3/2) (R_{\rm *, p}^{5}/G k_{\rm 2 *, p}) \sigma_{\rm *, p}$ \citep[cf. Eq.~78 in ][]{Eggletonetal98}. Note, however, that $\Delta t_{\rm *, p}$ depends on the radius of the body and its apsidal motion constant so, rigorously speaking, the constant time lag and the viscous time model are different}.

Tides inside the planet dominate the damping of the eccentricity. Specifically, considering a planet with $R_{\rm p} = 3 R_{\oplus}$ and $M_{\rm p} =10 M_{\oplus}$ in orbit around a sun-like star, we find:
\begin{equation}
\frac{K_{\rm p}}{K_{*}} = \frac{Q^{\prime}_{*}}{Q^{\prime}_{\rm p}} \left( \frac{M_{*}}{M_{\rm p}} \right)^{2} \left( \frac{R_{\rm p}}{R_{*}} \right)^{5}  \simeq 19 \frac{Q^{\prime}_{*}}{Q^{\prime}_{\rm p}}. 
\end{equation}
For a rocky planet, we can assume $Q^{\prime}_{\rm p} \sim 10^{2}$ by analogy with the Earth \citep{Rayetal96}, while for a gas giant $Q^{\prime}_{\rm p} \sim 10^{5}$ by analogy with Jupiter \citep{Laineyetal09}. On the other hand, for a solar-like star $Q^{\prime}_{*} \ga 10^{6}$ \citep{OgilvieLin07}, implying that the tidal dissipation inside the star is at least two orders of magnitude weaker than in the planet as far as eccentricity damping is concerned. { This result does not depend on the specific tidal model applied, but on the efficiency of the tidal dissipation inside the star and the planet, respectively. For example, for an orbital period of 10 days, the tidal lag time is $\Delta t_{*} \approx 3$~s for $Q^{\prime}_{*}=10^{6}$ for a solar-mass star. For a rocky planet the tidal lag time $\Delta t_{\rm p}$ is about $2-3$ orders of magnitude larger. 
Adopting the constant viscous time model, the dissipation constant is $\sigma_{*} \simeq 6.9 \times 10^{-55}$~kg$^{-1}$~s$^{-1}$~m$^{-2}$, while $\sigma_{\rm p} \simeq 4.4 \times 10^{-43}$~kg$^{-1}$~s$^{-1}$~m$^{-2}$ for a telluric planet. \citet{Hansen10} has further compared the different tidal dissipation parameterizations in systems consisting  of main-sequence stars and giant planets, so we refer the interested reader to that work.  }

When the eccentricity $e$ is close to the unity, $N_{e}(e) \gg \Omega_{e}(e)$ and with a little algebra we find:
\begin{eqnarray}
\label{dedt-time-lag}
\frac{1}{e} \frac{de}{dt}  & \simeq & -\frac{81}{2} \frac{1}{Q^{\prime}_{\rm p}} \left( \frac{M_{*}}{M_{\rm p}} \right) \left( \frac{R_{\rm p}}{a} \right)^{5} N_{e} (e)\, n  \\
 & \simeq & -3 \frac{1}{Q^{\prime}_{\rm p}} \left( \frac{M_{*}}{M_{\rm p}} \right) \left( \frac{R_{\rm p}}{a} \right)^{5} \frac{n}{(1-e)^{13/2}}. 
 \label{tidal_circ}
 \end{eqnarray}
Let us consider a planet with a semimajor axis $a=1$~AU in orbit around a sun-like star with $P_{\rm orb} = 1$~yr and a periastron distance of $0.05$~AU. The circularization of its orbit happens at virtually constant orbital angular momentum because most of the tidal dissipation occurs inside the planet whose rotation stays synchronized during the whole process. Therefore, the final orbital semimajor axis will be close to twice the periastron distance, i.e., $a_{\rm f} = a (1-e^{2}) \simeq 2 a (1-e) \sim 0.1$~AU for $e \sim 1$, corresponding to an orbital period of $\sim 11$~days. The circularization timescale, $\tau_{\rm c}^{-1} \equiv (1/e) (de/dt)$ for a rocky planet with $Q^{\prime}_{\rm p} \sim 10^{2}$, $R_{\rm p}= 2 R_{\oplus}$, and $M_{\rm p}= 4.2 M_{\oplus}$ as derived from Eq.~(\ref{tidal_circ}) is only $\approx 50$~Myr,  i.e., much shorter than the main-sequence lifetime of the system. Given the strong dependence on $(1-e)^{-13/2}$, $\tau_{\rm c}$ increases rapidly with the periastron distance, reaching $\sim 500$~Myr for 0.07~AU, i.e., a final orbital period of $\sim 20$~days. This limits  tidal circularization to final orbital periods below $\lesssim 15$~days in the case of a sun-like star. { These conclusions and the characteristic timescales of circularization are not significantly changed if we adopt the constant viscous time model with the above values of the dissipation constants $\sigma_{\rm *,p}$ because  the eccentricity evolution equation is  the same. }

On the other hand, large eccentricity excursions have a variety of durations, mostly ranging between $10^{5}$ and $10^{7}$~yr, owing to the chaotic nature of the orbital element fluctuations. Therefore, it is conceivable that during one of those excursions leading to a periastron distance $\la 0.05$~AU, the orbit of the planet can experience a strong tidal encounter with the star and be circularized \citep[cf. ][]{WuLithwick11}. This process does not significantly depend on the mass of the planet because for a more massive planet with a gaseous envelope the increase in $Q^{\prime}_{\rm p}$ is counterbalanced by the increase in $R_{\rm p}$ in Eq.~(\ref{tidal_circ}).  Therefore, the probability of tidal capture and the circularization timescale are not expect to vary remarkably with the mass of the planet. 

We can compare the above theoretical results with the measured eccentricity of the orbits of exoplanets as plotted, e.g., in Fig.~6 of \citet{UdrySantos07}. We see that  planets with an orbital period shorter than $\sim 6$~days, corresponding to a semimajor axis of $\sim 0.065$~AU, are on circular orbits, while a remarkable decrease of the eccentricity is observed for those having a period shorter than $\sim 20$~days that brings their periastron  closer than $0.06-0.08$~AU. This supports our conclusion that orbits with a periastron distance closer than $\approx 0.07$~AU are tidally circularized on a timescale shorter than the main-sequence lifetime of the host. Adding to Fig.~6 of \citet{UdrySantos07} recently discovered Earth-size planets, our conclusion is not changed because the eccentric orbits of some of them can be accounted for by the perturbations of close companions, given that most of them have been found to inhabit  multi-planet systems. On the other hand,  planets more massive that $\sim 20$~M$_{\oplus}$ show circular orbits below $\sim 0.065$~AU because they are often isolated or have distant companions.  

{ In principle, the eccentricity distributions should be different for candidate planets orbiting young and old stars, as defined in Sect.~\ref{Observations} on the basis of the detection of their rotational modulation. Specifically, older stars should display a more distant  limit for circularized orbits because tides have had more time to act. Unfortunately, a precise measurement of the orbital eccentricity is possible only through radial velocity measurements and is made difficult by the low masses of those planets that make their radial-velocity amplitudes of the order of a few m~s$^{-1}$. Therefore, we cannot perform such a detailed test of our model with our Kepler candidates, { but we can provide a prediction on the distribution of the orbital eccentricity. We recast Eq.~(\ref{dedt-time-lag}) in terms of the periastron distance $q \equiv a (1-e)$ using Kepler's third law as:
\begin{equation}
\frac{1}{e} \frac{de}{dt} \simeq -\frac{81}{2} \frac{1}{Q^{\prime}_{\rm p}} \left( \frac{M_{*}}{M_{\rm p}} \right) R_{\rm p}^{5}\,  N_{2}(e)\, q^{-13/2}, 
\end{equation}
where $N_{2}(e) = [1+(15/4)e^{2} + (15/8) e^{4} + (5/64) e^{6}]/(1+e)^{13/2}$ is maximum at $e=0$ and decreases for $e \rightarrow 1$ with a lower bound of $\sim 0.074$. During the tidal evolution, $q$ is bound between $q_{\rm i}$ and $2 q_{\rm i}$, where $q_{\rm i}$ is the value of the periastron distance when the tidal capture occurred, i.e., at the beginning of the evolution. Since 
$N_{2}$ is also bound, the timescale for orbital circularization $\tau_{\rm c}$ scales as $\tau_{\rm c} \propto q_{\rm i}^{13/2}$. The periastron distance in the case of an initially high eccentric orbit is $ q_{\rm i} \simeq a_{\rm f}/2$, where $a_{\rm f}$ is the final semimajor axis, i.e.,  when the orbit is circularized.  By applying Kepler's third law, we find: $P_{\rm c} (\tau_{\rm c}) \propto \tau_{\rm c}^{3/13}$, where $P_{\rm c}$ is the orbital period once it has been circularized after a tidal dissipation of duration $\tau_{\rm c}$. The longest possible $P_{\rm c}$ is obtained when the planet is injected into a highly eccentric orbit at the beginning of the main-sequence evolution of its host star and then is left unperturbed during all the subsequent tidal evolution. Therefore, we conclude  that the maximum $P_{\rm c}$ scales as $t^{3/13}$ with the mean age $t$ of the considered stellar population.  For systems with $P_{\rm orb} > P_{\rm c}$, we expect that the eccentricity values follow a Rayleigh distribution, as shown in Appendix~\ref{appendix}, because the efficiency of the tidal circularization drops rapidly with $q^{-13/2}$. Since the mean age difference in the case of the two populations introduced in Sect.~\ref{Observations}  does not exceed $\approx 50-100$ percent, the expected difference in $P_{\rm c}$ is less than $\approx 10-20$ percent because of the small exponent $3/13$ appearing in the power law dependence. In other words, we do not expect to see a remarkable difference in their circularization upper bound.  
 }


\subsubsection{Secular evolution of the orbit of the innermost planet}
\label{sec_evol}

With the theory introduced in Sect.~\ref{secular_chaos}, we can estimate the probability of getting a large orbital eccentricity for the innermost planet in a multi-planet system.  Its periastron distance will depend also on the initial semimajor axis of its orbit $a_{1}$. For the sake of simplicity, we assume that $a_{1}$ is the same for all the planets of the population we consider.  In the following, we show  that this assumption is adequate to account both for the correlation found by \citet{McQuillanetal13} and the other observations we  have discussed in Sect.~\ref{Observations} {pertaining to the planet and age distributions of stars with and without measured rotation periods.} 

For a population of systems with the same number of planets and initial semimajor axes, we consider a fixed value of the probability  $P_{\rm L}$ of having an eccentricity of the innermost planet  orbit $e \geq e_{0}$ at a given time $t$. The threshold value $e_{0}$ approaches the unity as the chaotic evolution goes on, according to Eq.~(\ref{prob_distr}). If the periastron distance gets closer than $\sim 0.05$~AU, tides can circularize the orbit and the final semimajor axis of the innermost planet becomes:
\begin{equation}
a_{\rm f} \simeq 2 a_{1} (1-e)  \leq  2a_{1} ( 1- e_{0}) = 2 a_{1} \left( \frac{2P_{\rm L}}{ | \varphi_{1}^{\prime}(1) | }\right)^{1/2}  t^{-1/2}, 
\label{afin}
\end{equation}
where  $t$ is the time elapsed since the stationary chaotic phase of the system was reached.  The orbital period of the innermost planet can be obtained from Kepler third law as a function of the initial orbital period $P_{\rm orb (i)}$ that corresponded to the initial semimajor axis $a_{1}$. We assume that the time $t$ is equal to the time elapsed since the beginning of the rotational evolution of the host star under the action of magnetic braking as parametrized by the gyrochronology relationship  in Eq.(3) of \citet{Barnes07} that we recast in the form:
\begin{equation}
\log t = \frac{1}{n_{\rm g}} (\log P_{\rm rot} - \log P_{\rm rot (i)} ),
\label{gyrochron}
\end{equation}
where $n_{\rm g} = 0.5189 \pm 0.007$, and the initial rotation period $\log P_{\rm rot (i)} \equiv \log a_{\rm g} + b_{\rm g}  \log [(B-V)- 0.4]$, with $a_{\rm g}= 0.7725 \pm 0.011$, $b_{\rm g}=0.601 \pm 0.024$, and $B-V \ga 0.45$ is the colour of the host star. Making use of Eq.~(\ref{gyrochron})  and Kepler third law, Eq.~(\ref{afin}) can be recast in the form:
\begin{equation}
\log P_{\rm rot} \leq -\frac{4n_{\rm g}}{3} \log P_{\rm orb} + c, 
\label{theor_corr}
\end{equation}
where the constant $c \equiv 2 n_{\rm g} \log [ 8 P_{\rm L}/ | \varphi_{1}^{\prime}(1)|]  +(4n_{\rm g}/3) \log P_{\rm orb (i)} + \log P_{\rm rot (i)}$ could be estimated if we knew $P_{\rm L}$, $\varphi_{1}^{\prime}(1)$, $P_{\rm orb(i)}$ and $P_{\rm rot(i)}$. 

The probability $P_{\rm L}$ could be estimated if we knew the frequency of the planetary systems that have an initial semimajor axis $a_{1}$, but for the moment we can treat it as a free parameter with the only limitation that $P_{\rm L} \ll 1$ for the validity of the linear approximation applied to derive Eq.~(\ref{prob_distr}). If we assume an initial rotation period $P_{\rm rot (i)} = 1$~day, independent of $B-V$, $a_{1} = 1$~AU, that corresponds to $P_{\rm orb (i)} = 1$~yr for a sun-like star,  $| \varphi_{1}^{\prime}(1) | = 0.081$ corresponding to $\sigma=0.25$ in Eq.~(\ref{varphi_value}), the value $c = 1.13 \pm 0.02 $ in Eq.~(\ref{mcquillan_corr}) gives $P_{\rm L} = (5.8 \pm 0.6) \times 10^{-4}$. If we adopt $\sigma = 0.35$ in Eq.~(\ref{varphi_value}), we obtain $P_{\rm L} = (7.1 \pm 0.7) \times 10^{-3}$. 
On the other hand, $4 n_{\rm g}/3 = 0.692 \pm 0.009$ is independent of all these assumptions,  and agrees closely with the slope of the linear regression by \citet{McQuillanetal13} in Eq.~(\ref{mcquillan_corr}). 

The above model was developed to explain the regression tracing the lower bound of the $P_{\rm rot}$-$P_{\rm orb}$ distribution in Fig.~\ref{protvsporb}. Nevertheless, we can test the validity of our assumptions by applying our model to the whole distribution. To warrant that the initial eccentric orbits produced by chaotic evolution be circularized, we restrict our comparison to the systems with a present orbital period $P_{\rm orb} \leq 15$~days, i.e., those whose periastron distance before circularization was $\sim 0.06$~AU. After a time $t$ since the beginning of the evolution, the probability of having an eccentricity in the interval $[e, e+de]$ in our linear approximation is: $P(e) de = -\varphi_{1}^{\prime}(1)\, (1-e)\, t\, de $. After tidal circularization, the final semimajor axis of the orbit will be:
\begin{equation}
a_{\rm f} \simeq 2 a_{1} (1-e) = 2 a_{1} \, \frac{P(e)}{\varphi_{1}^{\prime}(1)}\,  t^{-1}.
\label{af_prob}
\end{equation}
Since $a_{1}$ is assumed to be the same for all the considered systems, Eq.~(\ref{af_prob}) can be interpreted by saying that the probability of having a final semimajor axis in the interval $[a_{\rm f}, a_{\rm f} + da_{\rm f}]$ is proportional to $P(e)de$. By applying Kepler third law and the gyrochronology relationship, we  recast Eq.~(\ref{af_prob}) in the form:
\begin{equation}
\log [K P(e) ] = \frac{2}{3} \log P_{\rm orb} + \frac{1}{n_{\rm g}} \log P_{\rm rot},
\label{sys_prob}
\end{equation}
where 
\begin{equation}
K \equiv \frac{P_{\rm orb(i)}^{2/3} P_{\rm rot (i)}^{1/n_{\rm g}}}{| \varphi_{1}^{\prime} (1) |},
\end{equation}
is a constant in our model because we assumed that the initial orbital period and the initial rotation period are the same for all our stars. $K P(e)$ is proportional to the probability of having a system with an orbital and a rotation period in the ranges $[P_{\rm orb}, P_{\rm orb} + dP_{\rm orb}]$ and $[P_{\rm rot}, P_{\rm rot} + dP_{\rm rot}]$, respectively. Therefore, we expect to observe that the number $N$ of systems having a given value  of $K P(e)$ is proportional to the value of $K P(e)$ itself. 

We can test this prediction  by counting the observed number $N$ of systems in each bin of $K P(e)$. We plot the linear regression between $\log N$ and $\log [K P(e)]$ in Fig.~\ref{NKP}. The slope of the regression is $0.907 \pm 0.088$ as computed with the first eight points, i.e., excluding those with $\log K P (e)$ greater than 2.8. Since this is compatible with a unity slope within the uncertainties, we regard our model prediction as confirmed. 

For  $\log K P(e) \geq 3.1$, the number of observed systems decreases rapidly. The main reason is not the failure of our linear approximation to $P(e)$, but the increasing number of missed detections of rotation periods among stars with slow rotation. More precisely, since we have limited our comparison to the stars with $P_{\rm orb} \leq 15$~days, the values of $\log K P(e)$ larger than 3.1  correspond to a minimum rotation period of the host of $15$~days. Those stars are characterized by an increasing fraction of missed rotation  detections because for $P_{\rm rot} \gtrsim 20-25$~days sun-like stars have starspot lifetimes significantly shorter than the rotation period that makes the detection of their rotational modulation increasingly difficult (cf. Sect.~\ref{Observations}). On the other hand, the agreement between our simple model and the observations of stars with $P_{\rm rot} \lesssim 15-20$~ days is remarkably good. This suggests that indeed most of the short-period systems observed by Kepler are produced by the proposed mechanism, starting from an initially almost constant semimajor axis $a_{1}$.  Looking at the observed distribution of the orbital periods of giant planets in Fig.~4 of \citet{UdrySantos07}, we see that there is a broad peak between $\sim 300$ and $\sim 1400$ days. We may argue that those planets trace also the distribution of many Neptune- and Earth-sized planets that cannot be detected with the present techniques at those large orbital distances. In particular, the short-period component of the distribution peak, roughly centred around $\sim 1$~yr, might  represent the initial population giving rise to the short-period systems observed by Kepler. 

Given the detected frequency of short-period transiting planets in Kepler timeseries,  we can estimate the frequency of the systems forming  this initial population according to our model. 
Considering the 1251 candidates with $P_{\rm orb} \leq 15$~days and assuming a mean transit probability of $p_{\rm tr} \sim 0.09$\footnote{Corresponding to an orbit with $a=0.05$~AU around a sun-like star.}, we estimate a total of 13,900 short-period systems around the  150,000 stars observed by Kepler. Our estimated probability $P_{\rm L}$ for the lower 5 percent of the distribution implies an initial number of systems $N_{\rm in} \simeq 13,900 \times 0.05/P_{\rm L}$ about eight times the total number of stars observed by Kepler, if we adopt $\sigma= 0.25$ in the Rayleigh distribution of the eccentricities. This indicates that the value estimated for  $P_{\rm L}$ is underestimated by at least one order of magnitude for that value of $\sigma$ and supports our approach based on relative probabilities to test our model. On the other hand, if we assume $\sigma= 0.35$ in the eccentricity distribution, we obtain  that $N_{\rm in} \approx 10^{5}$ stars out of 150,000 observed by Kepler may be initially accompanied by a system of several Neptune- or Earth-sized planets with orbital periods $\gtrsim 1$~yr. { Note that the value of $\sigma$ and of the constant $c$ that appears in Eq.~(\ref{theor_corr}), together with the lower bound on $P_{\rm L}$ that comes from the maximum allowed $N_{\rm in}$, set an upper limit on $P_{\rm orb (i)}$ and therefore on the semimajor axis $a_{1}$ of the initial planet population considered in our simple model. More precisely, we find that a fixed value for $c$ implies $P_{\rm L} a_{\rm 1} \sim const$. With the adopted $\sigma = 0.35$, the upper bound for $a_{1}$ is $\sim 1.5$~AU in the case of a sun-like star. A lower bound can be found if one assumes that the frequency of multiplanet systems subject to the proposed chaotic evolution is, for example, at least $\approx 30$ percent for the stars of the Kepler sample. This gives $a_{1} \sim 0.5$~AU and $P_{\rm orb(i)} \sim 0.3$~yr. }

In principle, the problem of finding the most appropriate value of $\sigma$ can be settled by performing extensive numerical simulations on the evolution of planetary systems. However, such an approach is tremendously time-consuming \citep{Laskar08} and is hampered by the limited knowledge we have on the initial conditions. Therefore, in this first study we prefer an heuristic approach and adjust $\sigma$ to obtain acceptable results. 
\begin{figure}
\centering{
\includegraphics[width=6cm,height=8cm,angle=90]{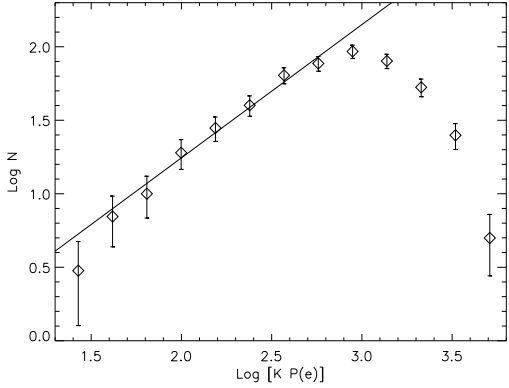}} 
\caption{Number of observed systems vs. the frequency predicted by our model for secular evolution. The errorbars correspond to the square root of the observed number of systems in each bin. The solid line is a linear regression computed with the eight data points with $\log K P(e) < 2.8$. Only systems with $P_{\rm orb} \leq 15$~days and $B-V \geq 0.5$ have been considered in order to have tidally circularized orbits and host stars that follow the gyrochronology relationship. }
\label{NKP}
\end{figure}

\subsubsection{Orbital period distribution}
\label{orb_period_distr}

We can compute the expected frequency of planetary systems from our model and compare it with the observations discussed in Sect.~\ref{Observations}.  We focus on the distribution of the 
orbital periods of the planets around stars with a measured rotation period and consider only systems with $P_{\rm orb} \leq 15$ days because tidal circularization is ineffective at longer periods.  The number $N$ of systems in a given orbital period bin  is proportional to the total probability $P$ for the systems belonging to that bin as estimated from Eq.~(\ref{sys_prob}). Therefore, we plot in Fig.~\ref{predicted_distr}, the ratio $N/KP$ vs. the orbital period for ten bins up to $P_{\rm orb} = 15$~days. We normalized the maximum of the ratio to the unity, given the uncertainty on the value of $K$. 
The prediction of a constant ratio is well verified for $3 \lesssim P_{\rm orb} \lesssim 15$~days, while the number of observed systems with $P_{\rm orb} \lesssim 3$~days around stars with measured rotation is 
remarkarbly smaller than predicted by our model. Since short-period planets are generally found around slowly rotating stars and only a small fraction of them has measured rotation, our result can be a consequence of the limited number of rotation detections in old stars. However, the discrepancy reaches about one order of magnitude in the case of the systems with $P_{\rm orb} \sim 0.5$~days, suggesting that a significant fraction of those planets have been removed by some mechanism not included in our simplified model. A good candidate is the tidal decay of the orbit
that we found to be significant for orbital period shorter than $2.5-3$ days (cf. Sect.~\ref{scen2}). This process can be accelerated by a restart of the chaotic evolution after the first orbital circularization, as we discuss in the next section. 
\begin{figure}
\centering{
\includegraphics[width=6cm,height=8cm,angle=90]{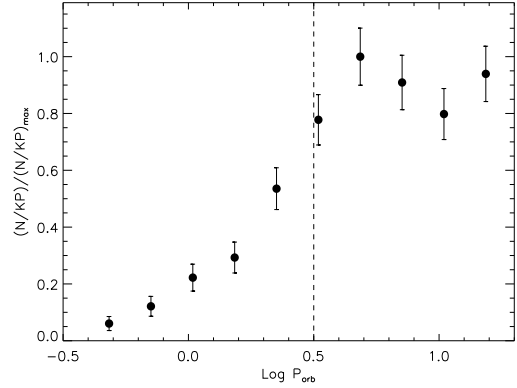}} 
\caption{Normalized ratio of the number of observed systems to the frequency predicted by our model vs. the orbital period. The errorbars correspond to the square root of the observed number of systems in each bin. Only systems with $P_{\rm orb} \leq 15$~days, a host with $B-V \geq 0.5$, and a measured host rotation period are considered. The vertical dashed line marks the minimum orbital period beyond which our model agrees with the observations. 
 }
\label{predicted_distr}
\end{figure}

\subsubsection{Late orbital evolution of planetary systems}

The dynamical evolution of the innermost planet did not end with the circularization of its orbit. This process dissipates its AMD, but the generally larger AMDs of the outer planets may restart a chaotic evolution making the orbit of the inner planet eccentric again.  However, given the larger separation between  the innermost planet and  the other  planets and the lower amount of AMD remained in the system, the eccentricity excitation  may require several Gyrs. In any case, its effect is that of making the  orbit of the innermost planet further approaching its host star until it reaches a minimum distance where precession and general relativity effects may stop the exchanges of AMD with the other planets. The closest final semimajor axis is \citep[cf., ][ Eq. 3]{WuLithwick11}:
\begin{equation}
a_{\rm f} \sim 6.4 \left( \frac{M_{\rm p}}{M_{\rm J}} \right)^{-1/5} \left( \frac{M_{\rm per}}{M_{\rm J}}\right)^{-1/5} \left( \frac{M_{*}}{M_{\odot}} \right)^{2/5} \left( \frac{\alpha}{1/6} \right)^{-3/5} \left( \frac{R_{\rm p}}{R_{\rm J}} \right) \; R_{\odot}, 
\label{a1fin}
\end{equation} 
where $M_{\rm p}$ and $R_{\rm p}$ are the mass and radius of the innermost planet, $M_{\rm per}$ the mass of the main perturbing planet, $R_{\rm J}$ the radius of Jupiter, and $\alpha \equiv a_{1}/a_{\rm per}$ the ratio of the semimajor axis of the innermost planet to that of its main perturber. For a hot Jupiter, this process can effectively set a minimum  semimajor axis, but for planets with a mass of a few Earth masses there is no such a limit because of their smaller radii \citep[cf. Sect.~5.1 of ][]{WuLithwick11}. 
At those close distances,  tidal dissipation {\it inside the star} becomes effective  and the planet begins to fall toward its host. A detailed discussion of this last phase is beyond the scope of the present work, so we do not consider the engulfment process here \citep[see, e.g.,][]{Jacksonetal09,Metzgeretal12}. We only limit ourselves to note that the final engulfment of close-in planets after several Gyrs of main-sequence evolution is in qualitative agreement with Fig.~\ref{porb_distr} where we see that the planet distribution  around main-sequence stars older than approximately $4-5$ Gyr is significantly depleted of short-period systems ($P_{\rm orb} \lesssim 15$~days) in comparison to that around younger stars.  Since most of those planets have radii $\lesssim 3$~R$_{\oplus}$, the halting mechanism suggested by \citet{WuLithwick11} is barely effective and they are allowed to approach  the Roche lobe limit after which they are rapidly engulfed. 
~\\

\section{Discussion and conclusions}

\noindent
We further explore the sample of 1961 KOIs considered by \citet{McQuillanetal13} in their search for rotational modulation in planetary hosts. We find that planetary candidates orbiting stars with measured rotation periods are more frequent  up to $P_{\rm orb} \sim 15$~days than candidates around stars with undetected rotation. By applying gyrochronology and considering  selection effects, we conclude that the sample of stars with measured rotation is likely to be younger than those without. Gyrochronology suggests an approximately uniform age distribution up to $\sim 3-4$~Gyr for the former sample, while the latter consists of stars with an age generally older than $4-5$~Gyr. 

We propose an explanation for the intriguing correlation found by \citet{McQuillanetal13} between the minimum orbital period of the innermost planet in a planetary system and the rotation period of its host star. Our model assumes that those planets come from a population with the same initial orbital semimajor axis that we tentatively assume to be $\approx 1$~AU in the case of a sun-like star. The secular orbital evolution of the innermost planet is driven by a process of chaotic diffusion of its eccentricity that can bring it to very high values. When this happens, the planet experiences a tidal encounter with its host star that circularizes and shrinks its orbit on a timescale of $\lesssim 50$~Myr. 
 By modeling the evolution of the eccentricity as a diffusion process, we are able to account for  the observed correlation between the rotation period of the host and the orbital period in the KOIs.  { Our model also predicts that  the eccentricity of the inner planet follows a Rayleigh distribution, except for orbital periods $P_{\rm orb} \leq P_{\rm c}$, where the circularization bound period depends on the mean age $t$ of the considered stellar population as $P_{\rm c} \propto t^{3/13}$.  This prediction can in principle be tested by future investigations. }
The proposed mechanism requires the presence of at least other $2-3$ distant planets in a typical system. This is in general agreement with the observed increase in the frequency of giant planets in the orbital period range between $\sim 300$ and $\sim 1400$ days \citep{UdrySantos07}, if we assume that those planets are accompanied by many smaller Neptune- and Earth-sized siblings that cannot be detected with current techniques. 

{ The present investigation indicates that  a simple model considering secular chaos and tidal evolution in planetary systems is consistent with the available observations, in particular for systems  consisting of Earth- or Neptune-sized planets detected through space-borne photometric monitoring. }
 For such systems, the detection of additional distant planets can be challenging with present techniques. Therefore, most of the apparently single planets discovered by Kepler could have distant companions with masses comparable to that of Neptune that cannot be presently detected, but can sustain the considered secular evolution. 
 
 We find that the distribution of the orbital periods of the planets around stars with measured rotation can be well reproduced by our simple model in the range $3 \lesssim P_{\rm orb} \lesssim  15$~days, suggesting that this interval of orbital periods is predominantly populated  by planets that were initially on much wider orbits and that had suffered the effects of secular chaos and of a final tidal circularization. For orbital periods longer than $\sim 15$ days, tidal circularization is ineffective  and a different mechanism must be invoked to populate this part of the orbital period distribution because secular chaos cannot modify the initial orbital period distribution. The KOI observations show a more or less constant frequency of planets up to  $P_{\rm orb} \sim 60$~days, suggesting that the mechanism should not critically depend on the finally reached orbital period. Migration in a protostellar disc during the pre-main-sequence evolution of the system may be a possible candidate \citep[e.g., ][]{Linetal96}.  

Secular chaos can effectively develop in systems similar to our own inner Solar System for which an excursion of the eccentricity of Mercury up to values close to the unity has been found in a few percent of the numerical evolutionary trajectories computed for a time span of $5$~Gyr \citep{Laskar08}. Therefore, it can contribute to bridge the gap  between our own solar system and extrasolar planetary systems in the theoretical investigation of their evolution. 

Our study also reveals the great potentiality of gyrochronology for a statistical analysis of the evolution of planetary system populations, suggesting that it is worth further investigating it, especially in the case of stars with close-in planets that could affect their magnetic braking process \citep[e.g., ][]{Lanza10,Brownetal11,PoppenhaegerWolk14}. 

\section*{Acknowledgments}

The authors are grateful to an anonymous Referee for a careful reading of the manuscript and several comments that helped to improve their presentation. A.F.L. is grateful to Drs.~C.~Damiani and A.~Vidotto for interesting discussions. 
The authors  gratefully acknowledge receipt of the solar total irradiance dataset (version d41\_62\_1302 in file ext\_composite\_d41\_62\_1302.dat of 10 March 2013) from PMOD/WRC, Davos, Switzerland,
and unpublished data from the VIRGO Experiment on the cooperative ESA/NASA Mission SoHO.  {E.S.~acknowledges support from NASA OSS Grant NNX13AH79G.}

\appendix

\appendix
\section[]{A random-walk diffusion model for the evolution of the eccentricity}
\label{appendix}

We consider the Poincar\'e variable $z$ of the orbit of the innermost planet \citep[cf. Sect.~2 in][]{LithwickWu11}:
\begin{equation}
z_{1} = \sqrt{2 \left( 1 - \sqrt{1 -e^{2}} \right)} \exp \left( \sqrt{-1}\, \omega \right), 
\end{equation}
where $e$ is the orbit eccentricity  and $\omega$ the argument of periastron. For $e \ll 1$ we can develop the above expression into a series of the eccentricity to obtain:
\begin{equation}
z_{1} \simeq  e \cos \omega + \sqrt{-1}\, e \sin \omega = k_{1} + \sqrt{-1} \, h_{1}, 
\end{equation}
where $k_{1}$ and $h_{1}$ are the Cartesian components of the Laplace vector introduced in Sect.~\ref{secular_chaos}.  According to \citet{LithwickWu11} and \citet{WuLithwick11}, the evolution of  $k_{1}$ and $h_{1}$  in a system consisting of several planets undergoing chaotic dynamical interactions can be approximately described as a random-walk diffusion process. We consider a simple model for such a process along the lines of the general theory of random walk processes. For simplicity sake, we refer to the variation of $k_{1}$, although the same considerations will be valid for $h_{1}$. We indicate with $\phi(s, t)\, \tau\, ds $ the probability of having a random variation of $k_{1}$, $\delta k_{1}$, in the interval $[s, s+ds]$ during the time interval $[t, t+\tau]$, where $\tau$ is the typical timescale among successive random-walk steps. Since $k_{1}$ is limited to the interval $]-1, 1[$, we have: 
\begin{eqnarray}
\int_{-1}^{1} \phi(s, t) ds & = & 1, \;\;\; \mbox{for all $t$, and} \nonumber \\
\phi(s,t) & = & \phi(-s,t), 
\label{eq3two}
\end{eqnarray}  
expressing the normalization of the probability distribution at any given time and the equal probability of positive and negative steps of the same length. We assume that $\phi$ depends on the time because the chaotic evolution becomes a stationary process only after some time interval of the order of hundreds of Myrs or a few Gyrs. 

The probability of having a value of $k_{1}$ in the interval $[k_{1}, k_{1}+dk_{1}]$ during the time interval $[t, t+dt]$ is given by  $\varphi_{k_{1}} (k_{1}, t) \, dk_{1} \, dt$, where $\varphi_{k_{1}}$ is the probability distribution function of $k_{1}$. Its variation in the short time interval $\tau$ can be expressed by a Taylor expansion truncated to the first order in the time:
\begin{equation}
\varphi_{k_{1}} ( k_{1}, t +\tau) = \varphi_{k_{1}} ( k_{1}, t ) + \tau \frac{\partial \varphi_{k_{1}}}{\partial t} + O(\tau^{2}), 
\label{eq5}
\end{equation}
where the partial derivative is evaluated in $(k_{1}, t)$. On the other hand, the change of the probability distribution obeys the equation:
\begin{equation}
\varphi_{k_{1}} ( k_{1}, t +\tau) = \int_{-1}^{1} \varphi(k_{1} -s, t) \phi(s, t) ds, 
\label{eq6}
\end{equation}
because any given value $k_{1}$ at the time $t+\tau$ comes from an initial value $k_{1} -s$ at the time $t$. We develop the r.h.s. of Eq.~(\ref{eq6}) into a Taylor series of powers of $s$ as:
\begin{eqnarray}
\varphi_{k_{1}} ( k_{1}, t +\tau) & = & \int_{-1}^{1}\varphi_{k_{1}} (k_{1}, t) \phi(s, t) ds  \nonumber \\
 & + &  \int_{-1}^{1} \frac{\partial \varphi_{k_{1}}}{\partial k_{1}} s \phi(s, t) ds \nonumber \\
 & + &  \int_{-1}^{1} \frac{\partial^{2} \varphi_{k_{1}}}{\partial k_{1}^{2}} s^{2} \phi(s, t) ds + O(s^{3}), 
 \label{eq7c}
\end{eqnarray}
where the partial derivatives are evaluated in the point $(k_{1}, t)$. 
The first and the second integrals on the r.h.s. of Eq.~(\ref{eq7c}) vanish because of the normalization of $\phi$ and its symmetry with respect to $s$ (cf. Eqs.~\ref{eq3two}). Neglecting terms of the third order or higher in $s$, we  obtain a diffusion equation for the probability distribution of $k_{1}$, i.e.:
\begin{equation}
\frac{\partial \varphi_{k_{1}}}{\partial t} = D(t) \frac{\partial^{2} \varphi_{k_{1}}}{\partial k_{1}^{2}}, 
\label{diffusion}
\end{equation}
where the diffusion coefficient
\begin{equation}
D(t) \equiv \frac{1}{2 \tau} \int_{-1}^{1} s^{2} \phi(s, t) ds,
\label{diffusion_coeff}
\end{equation}
depends on the time $t$. To solve Eq.~(\ref{diffusion}), we define a new variable $t^{\prime} \equiv \int D(t) dt$, so that it becomes:
\begin{equation}
\frac{\partial \varphi_{k_{1}}}{\partial t^{\prime}} = \frac{\partial^{2} \varphi_{k_{1}}}{\partial k_{1}^{2}}. 
\label{diffusion_prime}
\end{equation}
The solution of Eq.~(\ref{diffusion_prime}) verifying the boundary conditions $\varphi_{k_{1}}(\pm 1, t^{\prime})= 0$ can be well approximated for $k_{1} \ll 1$ as:
\begin{equation}
\varphi_{k_{1}}(k_{1}, t^{\prime}) = \frac{1}{\sqrt{4 \pi t^{\prime}}} \exp \left( -\frac{k_{1}^{2}}{4 t^{\prime}} \right).
\end{equation}
When the system reaches a stationary state, this distribution becomes independent of the time $t$. Mathematically, this is equivalent to say that $t^{\prime}$ becomes constant and we may put $t^{\prime} \equiv \sigma/2$ that gives a Gaussian distribution function for $k_{1}$ with a standard deviation $\sigma$. The same stationary distribution function is obtained for $h_{1}$. Therefore, the stationary distribution function for $e \simeq \sqrt{k_{1}^2 + h_{1}^{2}}$  is given by a Rayleigh distribution with standard deviation $\sigma$, in the limit $e \ll 1$. 

\bsp

\label{lastpage}

\end{document}